# Transforming the content, pedagogy and structure of an introductory physics course for life sciences majors

David P. Smith[1], Laurie E. McNeil, David T. Guynn, Alice D. Churukian, Duane L. Deardorff, Colin S. Wallace

*Department of Physics and Astronomy, University of North Carolina at Chapel Hill, Chapel Hill, NC, 27599*

**Abstract**

In this paper, we describe how we transformed our large-enrollment introductory physics sequence for life-science students to a Lecture/Studio format and aligned the physics concepts with authentic biological applications. We have reformed the pedagogy to include research-validated practices in interactive engagement, and accomplished our goals of enhanced learning gains, sustainability, and adoptability of our course reforms. The active engagement at the heart of the Lecture/Studio format results in comparable or enhanced learning gains (as measured by validated concept surveys) when compared to traditional instruction. When coupled with appropriate instructor preparation the format is sustainable, requiring no greater financial or human resources than does the traditional mode of teaching such courses. We have developed a complete suite of active-engagement instructional materials and made them available to the physics education community for adoption outside our institution.

## I. INTRODUCTION

One of the many instructional responsibilities of any physics department is to teach introductory physics to students who pursue majors in the life sciences. However, there is a concern in the physics education community (extending back several decades[2, 3] but growing significantly in the past fifteen years) that traditional physics courses taken by life sciences majors do not adequately elucidate how physics is useful in understanding the complex situations found in living systems. This concern has been spurred by multiple national reports[4, 5, 6] calling for transformations in these courses to address the problem that the basic physics concepts needed to build an understanding of biology are often absent from such traditional physics courses or may receive short shrift. To compound the problem, introductory physics for the life sciences (IPLS) courses are often taught in a traditional fashion that does not make use of the findings from physics education research (PER) regarding effective pedagogy. Consequently, life sciences students are often ill served by their physics instruction and are unable to use physics concepts effectively in their other studies and in their professional lives after graduation.

A number of physics departments across the nation have reacted to this situation by modifying the content of their IPLS courses.[7, 8, 9, 10, 11] Traditional topics of limited applicability to the life sciences (e.g., Kepler's laws, Gauss' law, AC circuits) have been replaced by topics that are critical for developing a physical understanding of the biological world but are not typically included in introductory physics courses for physical science majors (e.g., biological scaling, viscous fluid dynamics, diffusion). It is also common in the reformed courses to emphasize the role of physics concepts in understanding a biological system. For example, a coherent understanding of forces and torque is necessary for the understanding of animal locomotion, while the concept of electric potential is central to nerve signal propagation. These reforms to the content of IPLS courses have been accompanied by the development of suitable pedagogy supported by PER findings.[7]

Like all pedagogical reform efforts, these changes face the challenges of *sustainability* and *adoptability*. If life sciences students are to be better served by the physics courses they enroll in, changes to IPLS courses must persist once their designers and original instructors rotate to other instructional duties, and it must be feasible for other instructors in the same department and at other institutions to adopt the reforms.

The cross-disciplinary nature of these courses that makes them intellectually rich can also prove problematic for both sustainability and adoptability. Crouch and Heller[9] note that in order for an IPLS course to be sustainable it must lead to demonstrable improvements in student performance, be robust against sub-optimal implementations, be adoptable with reasonable effort, and also match instructors' beliefs related to teaching physics. They also note that it is critical to provide supporting materials to help the course instructors understand the aims and objectives of the new course design as well as the content contained in the biological contexts. O'Shea, Terry, and Benenson[10] also highlight the significant differences in the content and structure of a new IPLS course compared to that of a traditional physics course for physical science majors, and emphasize that the curriculum must be easily adoptable for successful implementation and sustainability. Examples exist of reformed courses that did not achieve these aims despite concerted effort, such as the one described by Meredith and Bolker[11] at the University of New Hampshire that was discontinued due to "diminishing resources, increased teaching requirements, and obstacles to granting workload credit to shared courses." Despite these difficulties, the sheer number of students who enroll in IPLS courses each year (in the hundreds at larger institutions) necessitates continued effort by the physics education community to serve these students well.

    At the University of North Carolina at Chapel Hill (UNC-CH) we took up this challenge and embarked on an ambitious project to improve student learning in our IPLS courses and enhance the ability of the students to apply physics concepts in biological contexts. We sought to do so in a way that could be sustained with the financial and human resources already available in the department and that could be adopted by other

institutions. To accomplish the transformation we assembled a team of faculty members with scholarly expertise in physics and astronomy education research, led by a senior professor familiar with PER findings but who did not have experience conducting such research. This team undertook a wholesale revision to the content, pedagogy, and structure of our two-semester IPLS course sequence, supported by a grant from the National Science Foundation.

Our content choices were based on the needs of life science majors, and we incorporated authentic biological contexts wherever possible. We structured the courses to use research-based active-engagement pedagogies. To assure sustainability and adoptability we made several strategic choices. First, we worked within our department's existing infrastructure so that the number of instructional personnel required to maintain the courses (faculty members and graduate and undergraduate student assistants) was the same in the original and transformed sequences. Second, we took the leap of transforming the entire multi-section course at once rather than beginning with a more limited pilot offering that could face pressure to revert to the old form in the face of the inevitable obstacles and minor failures. Third, we strove to make it possible for faculty members who did not participate in the initial reform process (and who have significant physics or astronomy research programs of their own) to prepare to teach the new course sequence with no more time and effort than they would expend in teaching any course for the first time. To accomplish this we created an entire course curriculum including lecture slides, in-class activities, reading assignments, homework assignments, and sample exam questions. We also utilized a team-teaching approach in which an instructor with experience teaching the course and with using active engagement pedagogies (the "mentor") co-teaches the course with a less-experienced colleague (the "apprentice"). These actions facilitate the adoption of the reformed course by instructors both within our department and at other institutions.

In this paper we describe the development of our reformed IPLS sequence and our strategies to ensure the sustainability and widespread adoption of our curriculum; we also present data supporting the effectiveness of the reforms for improved student learning. In section II we describe the context in which our transformation was carried out, including our student population and our interactions with colleagues in the life sciences. These interactions led us to the choices of course content that are described in that section. In section III we lay out the course structure we have adopted and its rationale. Section IV contains a description of the class activities we have designed (and made available for adoption outside of UNC-CH). In section V we discuss in detail how we prepare new course instructors. Finally in section VI we present an evaluation of our reforms.

## II. COURSE CONTEXT AND CONTENT

The two-semester IPLS sequence at UNC-CH has the largest enrollment of any physics courses taught by the department. Before the transformation, the ~600 students enrolled each semester were taught in a traditional lecture/laboratory format. Different lecture sections were often taught by faculty members who did not coordinate with one another, and the physics topics addressed in the laboratory sections were often disconnected from the presentation of the corresponding material in lecture. Of the students who enroll in this sequence, ~70% are majoring in the life sciences with the plurality (~42% of the total) being biology majors. 50-55% of them aspire to attend medical school and another 35-40% hope to pursue other health-science-related careers. We wanted to ensure that the new courses would contain content appropriate for students with these career goals.

Our goals for students and the content of the course were also defined through a series of conversations we had with faculty from multiple departments (including our own) who are stakeholders in the success of our course. We discovered that faculty from our department as well as from Biology, Environmental Science, Radiologic Science, and Psychology all had the same goals for IPLS students: they were passionate about their desire that life sciences students learn modes of thinking routinely employed in physics. All want the students to comprehend that physical principles can be expressed in mathematical terms and can be used to understand how the world works. Even more fundamentally, they would like the physics courses to help students learn to apply logic, formulate and apply simple models of natural processes, and draw inferences from simple measurements. Interestingly, although some faculty members were able to identify various specific concepts and phenomena that are important to biological systems (such as stress-strain relationships, optics relevant to instrumentation, and electrostatics), even among the physicists there was a perhaps-surprising lack of sentiment for "must-have" topics that no respectable physics course could leave out.

On the basis of these conversations, we felt quite free to choose to teach only those physics concepts that have authentic biological applications (or are foundational to the understanding of those concepts, such as the kinematics involved in the motion of jumping grasshoppers) and that would be accessible to both the students taking the course and the faculty in charge of the course. At the same time we eliminated those standard topics we deemed less relevant. For example, we chose to include diffusion and nonlinear stress and strain, but not planetary motion or rolling motion. Fig. 1 shows the topics that were originally taught in the first semester course, the topics now taught in the transformed course, and the topics common to both. Fig. 2 shows the same information for the second course in the sequence.

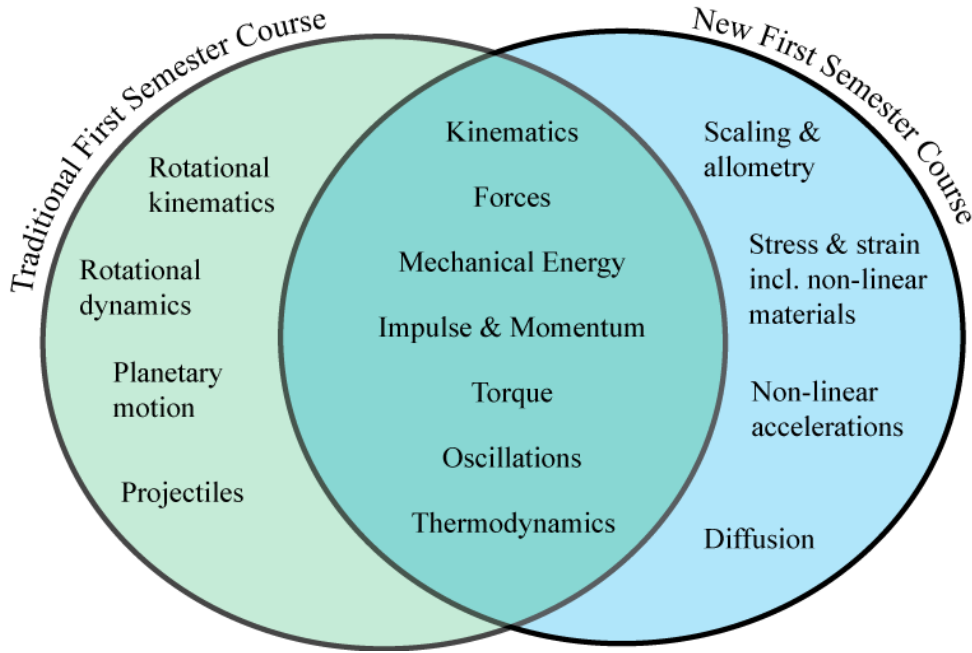

*Figure 1: Comparison of content in traditional and reformed first semester course.*

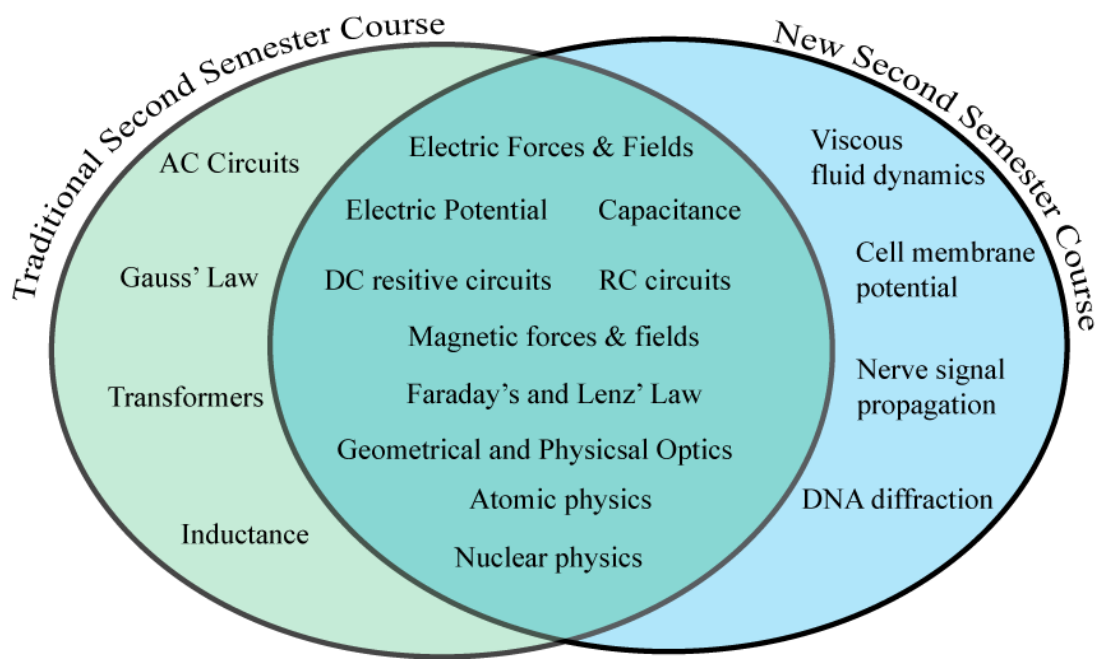

*Figure 2: Comparison of content in traditional and reformed second semester course.*

It is important to note that the teaching of the topics common to the traditional course was also transformed. For example, foundational principles such as Newton's laws and the conservation of energy were introduced in such a way to emphasize their importance in understanding a more complex biological system.

In addition to our choices of topics to include or eliminate, we made similar choices regarding the technical skills to emphasize in our instruction. In so doing we were mindful of the differences between the student population in the IPLS courses and that in our courses designed for physical science majors. For physics (and other physical science) majors, the introductory physics sequence is foundational, and these students will continue to build their physics understanding in subsequent courses. They will also encounter laboratory skills instruction primarily in a physics or chemistry context, where they will make use of error analysis and write laboratory reports in the manner expected in those disciplines. Students enrolled in IPLS courses at UNC-CH and elsewhere are likely to be in their junior or senior year of a biology major, and to have learned basic laboratory skills relevant to those disciplines in prior courses. However, their only opportunity to develop a robust understanding of the relevant physics concepts will be in these courses, as these are likely to be the only physics courses they take. Given the time limitations of a single two-semester sequence, we chose to focus on building our students' physics fluency as our primary goal. This meant that we had to place a lower emphasis on physics-specific laboratory skills, error analysis, and the writing of laboratory reports.

### III. REVISION OF COURSE STRUCTURE

While our choices of physics topics to include and to discard are similar to those made at other institutions, our decision to adopt the Lecture/Studio course structure that we describe below sets this work apart from transformations made elsewhere. We made this decision in order to accomplish our goal of implementing a sustainable program that led to improved student learning. While many research-validated pedagogies target specific components of the traditional lecture/laboratory/recitation structure of introductory physics courses (e.g., *Peer Instruction*[12] was originally conceived for use in lecture, *RealTime Physics*[13] for the laboratory, and the *Tutorials in Introductory Physics*[14] for recitation), there is also a long history of reimagining the course structure itself. These efforts to reform course structure often combine the lecture, laboratory, and recitation components into a single, unified instructional time period, with lecture de-emphasized in favor of collaborative group activities that focus on conceptual development, problem-solving skills, and/or taking and analyzing data. *Workshop Physics*,[15] developed at Dickinson College, and Studio Physics,[16] developed at the Rensselaer Polytechnic Institute, are examples of this kind of reimagined course structure for courses enrolling 20-60 students. The SCALE-UP model,[17] pioneered at North Carolina State University, demonstrated how the format of a Workshop or Studio Physics course could be replicated for a large-enrollment course serving over 100 students each semester.

In choosing a course structure we had to find a balance between producing large learning gains and having a sustainable model that fit within the department's infrastructure constraints. This led us to adopt the Lecture/Studio format (also known as New Studio), which was originally developed at Kansas State University[18] and which has also been adopted by the Colorado School of Mines.[19] The Lecture/Studio format is a hybrid of the traditional lecture/lab/recitation format and formats such as Workshop Physics, Studio Physics, and SCALE-UP. While the lecture component is retained and meets twice a week for 50 minutes at a time, laboratory and recitation are fused into a single 110-minute studio session that also meets twice a week. The Lecture/Studio model thus enables the majority of student contact time to be spent in the studio environment. The lectures and studio sessions alternate during the week, with a lecture and the studio session that follows it forming a coherent, tightly-linked module devoted to a particular topic. In the studio sessions students engage in a variety of active learning strategies, such as guided-inquiry laboratory investigations and pencil-and-paper activities inspired by both *Tutorials in Introductory Physics*[14] and the cooperative group problem solving material designed at the University of Minnesota.[20,21] A diagram of the weekly cycle is shown in Table 1 below.

*Table 1: Weekly schedule in the Lecture/Studio Model.*

| Monday | Tuesday | Wednesday | Thursday | Friday |
|---|---|---|---|---|
| Lectures (50 min each) | | Lectures (50 min each) | | Q & A session (optional) or exams during Lecture time |
| | Studios (110 min each) | | Studios (110 min each) | |
| Studios (110 min each) | | Studios (110 min each) | | |

The weekly cycle of class activities begins with an online assignment embodying *Just-in-Time Teaching* (JiTT):[22] students are required to complete a reading assignment (usually from the textbook) and answer questions on both the lecture and studio material and report what they find most puzzling. The instructor then uses the feedback to adjust his/her presentation accordingly for the Monday morning lecture. All students enrolled in the course attend an interactive lecture that utilizes pedagogical techniques such as *Peer Instruction*, *Ranking Tasks*,[23] and cooperative problem-solving to develop students' content fluency and to help prepare them for the upcoming studio. Three to four times each semester, 20 minutes at the beginning of a lecture is given to a faculty member from a life sciences department, who describes how the physics being studied applies to real biological contexts, including his/her own research. The studio sessions that follow each lecture are broken up into sections that typically hold 40-70 students (depending on the size of the classroom). During each studio session students work in groups of 3-4 students on one of the *Physics Activities for the Life Sciences* (*PALS*; see the following section), which extend and refine their understanding of the physics presented in the reading assignment and lecture as well as its biological relevance. The cycle then repeats with another module that begins with another pre-instruction JiTT assignment due before the Wednesday lecture and its subsequent studio session. Each module has an associated *MasteringPhysics*[24] online homework assignment that is typically due the following week. Friday mornings are reserved for an optional question-and-answer session and for midterm exams.

    The Lecture/Studio format successfully addresses many of the sustainability challenges that are associated with a studio-only format such as SCALE-UP. Since students are not all in studio at the same time, we do not need a single large studio room that can hold every student simultaneously. Instead we have converted our former introductory physics laboratory rooms into studio rooms that are used for multiple studio sections throughout the week. The number of teaching assistants (TAs) needed and the demands on their time are very similar to those of our prior lecture/laboratory format. For faculty members, teaching these classes does require more time and effort than would an upper-division course the faculty member has taught before, especially since the department strongly encourages faculty members to teach a studio section as well as deliver lectures (this is a requirement for all first-time instructors, as described in more detail in Section V). This burden can be reduced by offering faculty members who teach these courses for several semesters a reduced teaching load in a subsequent semester. Our department has judged the Lecture/Studio format to be sustainable for the foreseeable future, and all introductory physics courses at UNC-CH, including both the IPLS sequence and the sequence for physical science majors (which was transformed into this format earlier), are now only offered as Lecture/Studio courses.

## IV. PHYSICS ACTIVITIES FOR THE LIFE SCIENCES (PALS)

In developing the modules that make up these courses we have developed more than fifty studio activities (called *Physics Activities for the Life Sciences*, or *PALS*), combining existing ideas and activities (properly attributed) with our own newly-created ones. The activities address important physical principles and their applications to the life sciences. Many of them focus on topics that are not part of the traditional introductory physics curriculum (see Fig. 1 and Fig. 2), including stress and strain, diffusion, chemical energy, and life at low Reynolds numbers.[25] Each module focuses on a specific topic, although some fundamental topics (e.g., Newton's laws and electric potential) require multiple modules. In addition to emphasizing the applications of physics to the life sciences, we designed the *PALS* to address common student conceptual and problem-solving difficulties, as identified in the PER literature.[26, 27, 28, 29, 30, 31, 32, 33, 34, 35, 36, 37, 38, 39] For some topics, such as diffusion, osmosis, and life at low Reynolds numbers, there was a paucity of PER literature to draw upon. These activities thus represent topics for which future physics education research may be conducted.

The *PALS* comprise a mixture of tutorial-like and laboratory-like activities, with some *PALS* being purely pencil-and-paper activities, some being laboratory investigations, and some being a mix of the two. The pedagogy of the *PALS* was inspired by research-validated and research-supported activities, such as the *Tutorials in Introductory Physics*, *Tasks Inspired by Physics Education Research* (TIPERs),[40,41] and *Ranking Tasks*. In some cases we modified an activity developed elsewhere (giving proper attribution to its original developers), such as the Breathing Worms activity developed at the University of Maryland,[42] the Nerve Signals activity developed at Swarthmore College,[43] and the DNA Diffraction Pattern activity developed at the University of Wisconsin-Madison.[44] Some *PALS* utilize PhET simulations.[45] Simple computational work is also included throughout the *PALS* curriculum, primarily focused on numerical integration techniques and linear, log-log, and semi-log graphing and trend-fitting. The laboratory investigations in the *PALS* focus on getting students to make predictions about the result of a particular experiment and then perform the experiment in order to check whether their predictions were correct. In this respect, some *PALS* employ the elicit-confront-resolve methodology[46] successfully used in other curricula. Other *PALS* build upon and refine students' pre-existing intuitions, which has also been shown to be a successful pedagogical methodology.[47, 48] The *PALS* are collaborative activities meant to be completed by groups of three to four students working together in the studio environment. Each activity has been appropriately scaled such that it can be completed by the average group within a single 110-minute studio period.

The *PALS* constitute the core of our transformed courses, and developing them has helped make the transformation sustainable even when the development team has moved on to other projects. Faculty members who teach one of the transformed IPLS courses do not have to develop any studio activities and can concentrate on effective delivery of the content of the modules. The creation of these studio activities has also allowed us to clearly define the learning goals for each module, goals that govern the content and structure of the accompanying lectures, homework, and exam questions. This inclusive approach to the development process has also made it easier for us to offer our instructional materials for adoption at other institutions. The *PALS* are available via comPADRE, and the ancillary materials (lecture slides, reading assignments, JiTT materials, sample exam questions) are available upon request.

## V. INSTRUCTOR PREPARATION

Another element contributing to the sustainability of these courses is the process we have instituted for preparing instructors to teach these courses. While decades of research both within physics and across disciplines have clearly demonstrated that active learning environments are necessary to achieve high student learning gains,[49] the use of active learning pedagogies is not enough to guarantee significant increases in student understanding. An instructor's effectiveness at implementing active learning plays a critical role in determining how much his/her students learn.[50] Additionally, research has shown that many instructors who adopt active learning techniques modify them in ways that may attenuate their effectiveness, and may eventually abandon these techniques altogether and return to traditional lecture as the sole method of instruction.[51] In order to ensure that the curriculum developed for the transformed IPLS courses (including the *PALS*) is implemented within our department as intended and sustained regardless of who is assigned to teach it, we initiated several new instructor preparation procedures.

Each of the transformed courses is co-taught by two or more faculty members each semester. As noted in Section II, the IPLS courses at UNC-CH have the largest enrollments of any physics courses offered by the Department of Physics and Astronomy. With 250-400 students enrolling in a single IPLS course, we typically split the class into two lecture sections that are held back-to-back in the same room and with roughly equal numbers of students. (While it would be possible to operate the transformed course with a single lecture section, the lack of available large lecture rooms and considerations for student class scheduling led us to use two sections.) Prior to our transformation efforts, each faculty member assigned to teach the course would assume complete responsibility for one of the lecture sections. It was common for the instructors to move at different paces, emphasize different topics, and create their own homework and exams. Post-transformation, the instructors now share the responsibilities for both sections. The instructors decide who will give the back-to-back lectures for a given module, and all instructors are present for the lectures and assist in the activities that take place during it. Other duties are likewise split, with one instructor serving as the official instructor-of-record who is the final arbiter of student grades, while another may be the "studio coordinator" and oversee all aspects of the course relating to the studios. However, major decisions are made jointly, even though one of the instructors may be designated as the person to communicate those decisions to the students. The instructors meet once per week to discuss the progress of the course, plan exams, and coordinate their responses to any unique situations involving individual students. This model of co-teaching the courses reduces the administrative burden placed on any individual faculty member involved with the course.

    The first time a faculty member teaches a course, he or she is paired with a co-instructor who has experience teaching the course. We refer to the first-time instructor as the "apprentice" and the experienced instructor as the "mentor." Note that the designation of "mentor" or "apprentice" does not necessarily correlate with faculty rank; a tenured full professor may be the apprentice to a much more junior non-tenure-track instructor serving as mentor. The apprentice must lead one of the studio sections (the

remainder are taught by graduate and undergraduate assistants), which allows the apprentice to become familiar with the *PALS* and with how studios are taught.  This is vital since the content of the *PALS* is the primary driver of the curriculum and students spend 2/3 of their class time in studio.  The apprentice also delivers some of the lectures (typically one-third to one-half of the total).  Over the course of the semester the mentor provides feedback to the apprentice on his or her implementation of the active learning pedagogies embedded in the curriculum.  This long-term, real-time, continuous feedback between the mentor and the apprentice is consistent with the best practices for effectively supporting instructors to adopt active learning pedagogies.[52]  After a semester of mentored experience in the course, the apprentice can become a mentor to an inexperienced faculty member, thereby continuing to expand the cohort of faculty members who have used active-engagement pedagogy.  This can also lead to faculty members employing such techniques in other, more advanced courses that they teach.

    The full instructional team for a course also includes the graduate and undergraduate students who serve as Teaching Assistants and Learning Assistants in the studio sections.  All members of the team assemble each Friday for a two-hour meeting devoted to preparing for the coming week's studios.  During these Friday meetings the team works through each question on the upcoming studios collaboratively, just as we expect students to do.  This ensures that each instructor understands and agrees upon the answers to each question, understands the procedures and equipment (if any) to be used, and is aware of common student conceptual, reasoning, and problem-solving difficulties as well as strategies to help students overcome those difficulties.  These meetings are essential to foster communication among all members of the instructional team and to ensure that

everyone understands the goals for students and their roles in helping students achieve those goals. This immersion in active-engagement pedagogy and the rationale behind it has also had the salutary effect of making it more likely that faculty members (and graduate students who later become faculty members) adopt such methods in other courses they teach, a phenomenon we have begun to observe in our department.

## VI. COURSE EVALUATION

As we have described above, our goals for the sustainability of our course transformation have been met to our satisfaction. To assess our success in achieving our goals for enhanced student learning we needed to measure that learning. The primary method of data collection we have used in this transformation project is written concept surveys. For our first-semester sequence (which focuses primarily on mechanics) we administered the Force Concept Inventory (FCI),[53] and for the second semester sequence (which emphasizes electricity and magnetism) we used the Conceptual Survey of Electricity and Magnetism (CSEM).[54] We began collecting these data well before we began the transformation process and therefore have a comprehensive before-and-after picture of how our learning gains were affected by the transformation. These concept inventories cover only a fraction of the content we teach, but they have the virtue of being validated and nationally normed.

Our analysis used the average normalized gain[55] (commonly known as the "Hake gain") to quantify learning gains on the FCI and the CSEM. To help track the changes in student performance in the first semester course, we have split the data into three categories according to the teaching style and pedagogy used in the courses. For semesters Fall 2008 to Fall 2012, teaching styles in the first-semester course were close to traditional methods of lecturing with some use of interactive engagement strategies by some faculty. In Spring 2013, extensive use of *Peer Instruction* tasks as well as materials from *Tutorials in Introductory Physics* were introduced into the lectures for the first course in the sequence. We would classify the course in this period as highly interactive, but the content and structure was still quite traditional. The Lecture/Studio format and revised content were introduced in Fall 2014. FCI data were collected beginning in Fall 2008 and a graph of the average normalized gain for each semester through Spring 2017 is shown in Fig. 3. The three categories of pedagogy (traditional structure/traditional presentation, traditional structure/active presentation, new structure/active presentation) are distinguished by the color and shading of the bars in the figure.

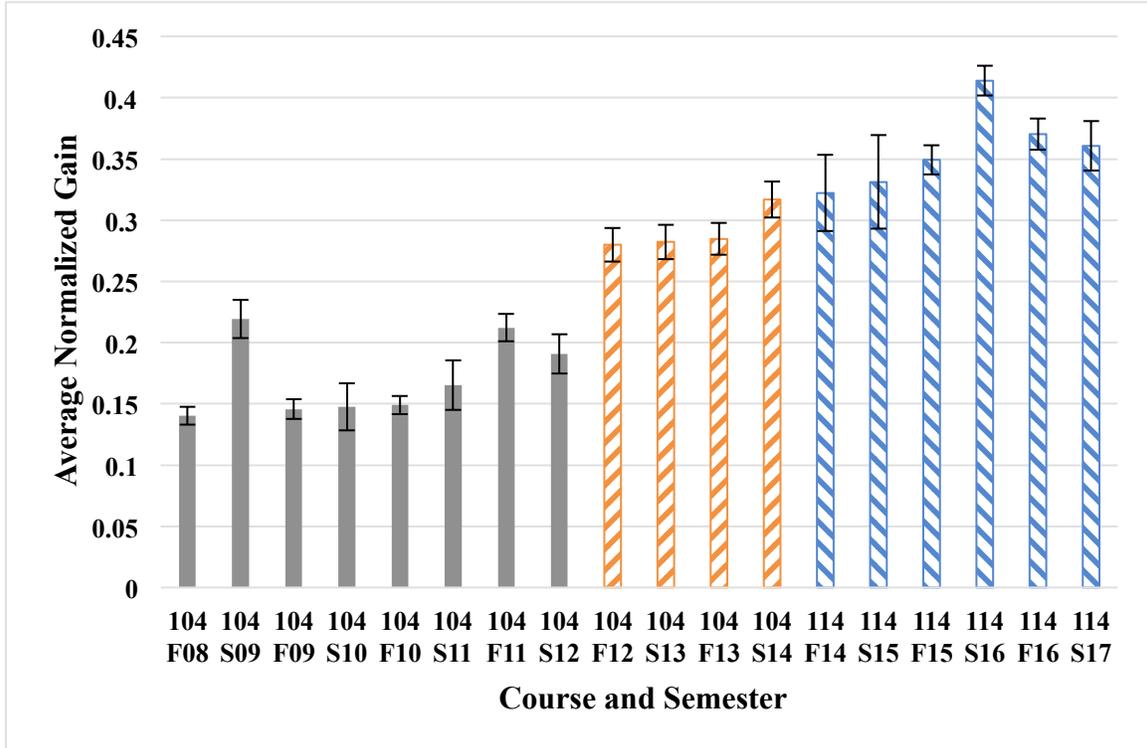

*Figure 3: Hake Gain for the Force Concept Inventory. PHYS 104 was the pre-transformed version of the first semester IPLS course at UNC-Chapel Hill; PHYS 114 is the transformed version. Solid grey bars are courses taught traditionally, orange bars with upward-slanting hash marks are courses taught in traditional structure and content but with some active engagement in lecture, and blue bars with downward-slanting hash marks are courses taught in Lecture/Studio mode.. Error bars represent standard error.*

The graph shows an overall upward trend in the FCI scores as the course pedagogy was changed from traditional methods to interactive engagement methods. Beginning in Spring 2014, the average normalized gain reached values above 0.3, which is defined by Hake as moderate-gain.[55]

Analysis of the FCI data show that students in the new course sequence perform at least at the same level as an interactive engagement class teaching traditional content, i.e. that the biological focus has not led to a decrease in understanding of fundamental physics concepts. In Spring 2016 (the fourth semester the reformed class was taught) the highest FCI average normalized gains ever recorded for the course to date were achieved.

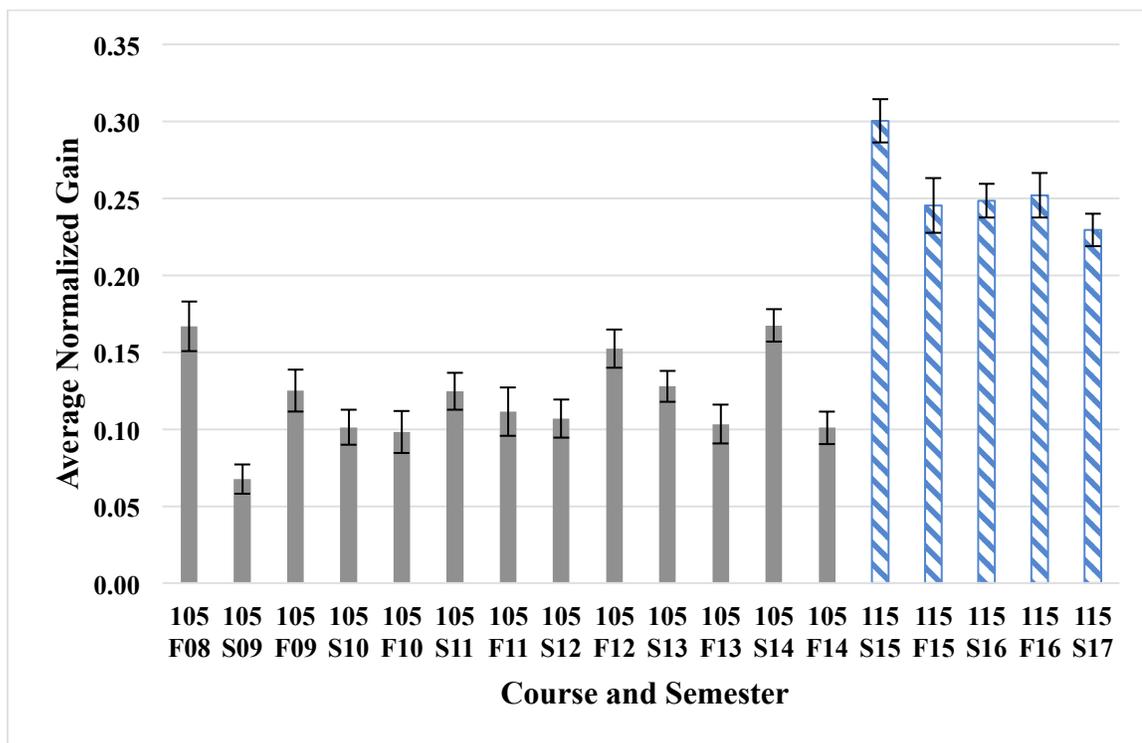

*Figure 4: Hake Gain for the CSEM. PHYS 105 was the pre-transformed version of the first semester IPLS course at UNC-Chapel Hill; PHYS 115 is the transformed version. Solid grey bars are courses taught traditionally and blue bars with hash marks are courses taught in Lecture/Studio mode. Error bars represent standard error.*

The results for the learning gains achieved in the second-semester course, assessed using the normalized gain on the CSEM, are shown in Figure 4. The mode of instruction for this course was not changed prior to the transition to the Lecture/Studio format (including the new course content) in Spring 2015. The measured learning gains for the courses taught in the Lecture/Studio format were strikingly larger than any measured gain for the course in its traditional format. Furthermore, while average normalized gains between 0.20 and 0.30 may seem low based on the scale proposed by Hake,[55] one must keep in mind that students generally find the CSEM to be a more difficult assessment than the FCI. For example, Maloney *et al*. find that introductory algebra-based physics students have average pre- and post-test CSEM scores of 25% and 44%, respectively, while introductory calculus-based physics students have pre- and post-test scores of 31% and 47%, respectively. These correspond to average normalized gains of 0.25 for the algebra-based physics students and 0.23 for the calculus-based physics students, which, as Fig. 4 shows, are comparable to the results obtained by students in the transformed IPLS course. These findings indicate that the format in which the course is taught has a statistically significant impact on learning gains as measured by the CSEM.

Based on our concept survey results we conclude that changing to active-engagement pedagogy has had a noticeable positive influence on the learning gains achieved by our students. Further, our emphasis on biological relevance has not reduced the degree to which our students grasp the fundamental physics concepts measured by these concept surveys.

## VII. CONCLUSION

In this project, we have developed a complete suite of active-engagement instructional materials that incorporates research-validated practices and enables a two-semester introductory physics sequence for life sciences students to be taught in a large-enrollment Lecture/Studio format. When coupled with appropriate instructor preparation, the format is sustainable in that it requires no greater financial or human resources than does the traditional mode of teaching such courses, and can be maintained intact as different instructors rotate in and out of the courses. The active engagement at the heart of the Lecture/Studio format results in comparable or enhanced learning gains as measured by validated concept surveys. The revised curriculum and course format also provides students with many opportunities to apply physics concepts in biological contexts. We have prepared our instructional materials in a form amenable to adoption and made them freely available to the physics education community. By doing so we hope that we will help the number of physics departments offering such courses to increase, improving the education of large numbers of life-science students.

ACKNOWLEDGEMENTS

The work described in this paper was carried out with the support of the National Science Foundation under grant number DUE-1323008. Earlier reforms that this effort built upon were supported by NSF under grant number DUE-0511128.